\documentclass[12pt]{article}
\usepackage{authblk}
\usepackage{graphicx} 
\usepackage{amsmath,amssymb,amsthm}
\usepackage{dsfont}
\usepackage{indentfirst}
\usepackage{color}
\usepackage{xcolor}
\usepackage{algorithm}
\usepackage{algpseudocode}
\usepackage{hyperref}
\usepackage{lineno} 
\usepackage[margin=1in]{geometry}
\usepackage[authoryear]{natbib}
\usepackage[toc,page]{appendix}
\usepackage{subcaption}
\usepackage[rightcaption]{sidecap}
\usepackage{wrapfig}
\usepackage{setspace}

\title{A Bayesian Mechanistic Model of Spatiotemporal Dynamics in Invasive Tree Species}
\author[1,2]{Jiaqi Chen\thanks{Email: \texttt{jchen61@huskers.unl.edu}}}
\author[3]{Huijing Du\thanks{Email: \texttt{hdu@unl.edu}}}
\author[4]{Yawen Guan\thanks{Corresponding author. Email: \texttt{yawen.guan@colostate.edu}}}
\affil[1]{Department of Statistics, University of Nebraska--Lincoln}
\affil[1]{Department of Radiation Oncology, University of Nebraska Medical Center}
\affil[3]{Department of Mathematics, University of Nebraska--Lincoln}
\affil[4]{Department of Statistics, Colorado State University}
\date{}

\begin{document}
\newcommand{\yg}[1]{\textcolor{purple}{(YG: #1)}}

\newcommand{\btheta}{ \mbox{\boldmath $\theta$}}
\newcommand{\bmu}{ \mbox{\boldmath $\mu$}}
\newcommand{\balpha}{ \mbox{\boldmath $\alpha$}}
\newcommand{\bbeta}{ \mbox{\boldmath $\beta$}}
\newcommand{\bmeta}{ \mbox{\boldmath $\eta$}}
\newcommand{\bdelta}{ \mbox{\boldmath $\delta$}}
\newcommand{\blambda}{ \mbox{\boldmath $\lambda$}}
\newcommand{\bgamma}{ \mbox{\boldmath $\gamma$}}
\newcommand{\brho}{ \mbox{\boldmath $\rho$}}
\newcommand{\bpsi}{ \mbox{\boldmath $\psi$}}
\newcommand{\bepsilon}{ \mbox{\boldmath $\epsilon$}}
\newcommand{\bomega}{ \mbox{\boldmath $\omega$}}
\newcommand{\bDelta}{ \mbox{\boldmath $\Delta$}}
\newcommand{\bSigma}{ \mbox{\boldmath $\Sigma$}}
\newcommand{\bOmega}{ \mbox{\boldmath $\Omega$}}
\newcommand{\bGamma}{ \mbox{\boldmath $\Gamma$}}

\newcommand{\bA}{ \mbox{\bf A}}
\newcommand{\ba}{ \mbox{\bf a}}
\newcommand{\be}{ \mbox{\bf e}}
\newcommand{\bP}{ \mbox{\bf P}}
\newcommand{\bx}{ \mbox{\bf x}}
\newcommand{\bX}{ \mbox{\bf X}}
\newcommand{\bB}{ \mbox{\bf B}}
\newcommand{\bZ}{ \mbox{\bf Z}}
\newcommand{\by}{ \mbox{\bf y}}
\newcommand{\bY}{ \mbox{\bf Y}}
\newcommand{\bz}{ \mbox{\bf z}}
\newcommand{\bh}{ \mbox{\bf h}}
\newcommand{\bmm}{ \mbox{\bf m}}
\newcommand{\br}{ \mbox{\bf r}}
\newcommand{\bt}{ \mbox{\bf t}}
\newcommand{\bs}{ \mbox{\bf s}}
\newcommand{\bb}{ \mbox{\bf b}}
\newcommand{\bff}{ \mbox{\bf f}}
\newcommand{\bL}{ \mbox{\bf L}}
\newcommand{\bU}{ \mbox{\bf U}}
\newcommand{\bu}{ \mbox{\bf u}}
\newcommand{\bv}{ \mbox{\bf v}}
\newcommand{\bV}{ \mbox{\bf V}}
\newcommand{\bG}{ \mbox{\bf G}}
\newcommand{\bC}{ \mbox{\bf C}}
\newcommand{\bg}{ \mbox{\bf g}}
\newcommand{\bH}{ \mbox{\bf H}}
\newcommand{\bI}{ \mbox{\bf I}}
\newcommand{\bD}{ \mbox{\bf D}}
\newcommand{\bM}{ \mbox{\bf M}}
\newcommand{\bW}{ \mbox{\bf W}}
\newcommand{\bw}{ \mbox{\bf w}}
\newcommand{\bT}{ \mbox{\bf T}}
\newcommand{\bQ}{ \mbox{\bf Q}}
\newcommand{\bR}{ \mbox{\bf R}}
\newcommand{\bfe}{ \mbox{\bf e}}
\newcommand{\bzero}{ \mbox{\bf 0}}

\newcommand{\iid}{\stackrel{iid}{\sim}}
\newcommand{\indep}{\stackrel{indep}{\sim}}
\newcommand{\Xtil}{{\tilde X}}
\newcommand{\bXtil}{\bf {\tilde X}}

\newcommand{\calR}{{\cal R}}
\newcommand{\calL}{{\cal L}}
\newcommand{\calG}{{\cal G}}
\newcommand{\calD}{{\cal D}}
\newcommand{\calC}{{\cal C}}
\newcommand{\calS}{{\cal S}}
\newcommand{\calB}{{\cal B}}
\newcommand{\calA}{{\cal A}}
\newcommand{\calT}{{\cal T}}
\newcommand{\calO}{{\cal O}}
\newcommand{\calN}{{\cal N}}
\newcommand{\calK}{{\cal K}}

\newcommand{\pd}{\partial}

\newcommand{\argmax}{{\mathop{\rm arg\, max}}}
\newcommand{\argmin}{{\mathop{\rm arg\, min}}}
\newcommand{\Frechet}{\mbox{Fr$\acute{\mbox{e}}$chet}}
\newcommand{\Matern}{ \mbox{Mat$\acute{\mbox{e}}$rn}}
\newcommand{\Prob}{\mbox{Prob}}

\newcommand{\beq}{ \begin{equation}}
\newcommand{\eeq}{ \end{equation}}
\newcommand{\beqn}{ \begin{eqnarray}}
\newcommand{\eeqn}{ \end{eqnarray}}

\begin{spacing}{1}
\maketitle
\end{spacing}

\begin{abstract}
\begin{spacing}{1}
Modeling the spatiotemporal dynamics of invasive species requires frameworks that represent spread and growth mechanisms explicitly while accommodating the distributional features of ecological data. We propose a Bayesian mechanistic framework for invasive tree cover that combines an ecological diffusion--logistic growth partial differential equation (PDE) with a zero-inflated binomial observation model. The diffusion and growth coefficients are modeled as functions of environmental covariates, enabling direct inference on the ecological processes driving encroachment, and the zero-inflation component separates structural absence from low observed cover. To enable scalable Bayesian inference on satellite-resolution grids, we derive a homogenized equation for ecological diffusion with spatially varying diffusion and growth coefficients, which reduces computation while preserving large-scale spatial structure. Applied to satellite-derived tree cover in a subregion of the Nebraska Sandhills, where eastern redcedar is encroaching on grassland, the model indicates that elevation is strongly associated with where tree cover accumulates but only weakly with its local growth rate, whereas distance to the nearest stream is associated with both. Compared with a generalized additive model, a spatial hierarchical model fitted via INLA, and a regression tree, the proposed method directly estimates these processes as interpretable functions of environmental covariates and gives the best-calibrated predictive uncertainty.

\vspace{1em}
\noindent\textbf{Keywords:} Ecological diffusion; Zero-inflation; Mechanistic model; Partial differential equation; Homogenization. 
\end{spacing}
\end{abstract}

\section{Introduction}
Invasive tree species can establish and spread across diverse landscapes \citep{vanWilgen2001}. In the grasslands of the central United States, the dominant encroaching species is eastern redcedar (\emph{Juniperus virginiana}; ERC), which has expanded across Kansas, Oklahoma, Missouri, and Nebraska \citep{meneguzzo_status_2015}. The shift from grassland to redcedar woodland reduces endemic biodiversity and lowers the profitability of grazing land \citep{briggs_assessing_2002, Symstad&Leis2017}. The Nebraska Sandhills, one of the largest intact prairie ecosystems in North America, are among the regions where this encroachment is under way. 

Managing encroachment requires knowing where and how fast ERC spreads, yet such information is scarce. ERC observations come mainly from field surveys, while historical records with wide spatial coverage are lacking \citep{meneguzzo_status_2015}. Satellite-derived tree cover offers an alternative. The Rangeland Analysis Platform provides annual 30~m estimates from Landsat imagery back to 1985 \citep{landsat_2021, Allred2021}. Because the Sandhills are predominantly grassland, rapid increases in tree cover there can largely be attributed to ERC \citep{shrestha_current_2025}, making these data a proxy for its spread.




Mechanistic models offer a natural way to study the processes behind such spread. Diffusion–growth PDEs have been used to describe invasive tree spread \citep{efendiev_existence_2026} and tree--grass dynamics \citep{Yatat2018}, but their coefficients were not estimated from data. Deterministic PDEs have been embedded in hierarchical Bayesian models to estimate the spread of invasive animals \citep{Hooten2007,Williams2017} and disease \citep{Garlick2011,Hefley2017}, whereas stochastic PDEs suit settings where process-level stochasticity plays a dominant role \citep{Hanks2017, Wikle2022}. Year-to-year changes in tree cover are small relative to the noise in its satellite-derived estimates. We therefore adopt a deterministic PDE and place the uncertainty in a flexible observation model.

Two features of the tree cover data pose methodological challenges for fitting a mechanistic PDE within a Bayesian hierarchical model. First, the data are recorded at 30~m resolution, which yields 90,000 pixels per year even for the small subregion studied here, so solving the PDE numerically at every MCMC iteration is computationally prohibitive. Homogenization reduces this cost by solving an averaged equation on a coarse grid while retaining the influence of fine-scale heterogeneity \citep{Holmes2013, Pavliotis&Stuart2008}. Homogenized equations have been derived for ecological diffusion with exponential growth \citep{Garlick2011,Hefley2017a}, and with logistic growth when the growth rate is spatially constant \citep{Lu2019}, but not when both the diffusion and growth coefficients vary in space.
 
Second, tree cover is recorded as an integer percentage, with many zeros in early years and some pixels at complete (100\%) cover in later years. Standard models handle these features only partially or indirectly. The Beta distribution places no mass at either boundary, and its zero- or one-inflated extensions \citep{Ospina2012} treat the data as continuous and require a separate mixture component for each boundary. The zero-inflated Poisson \citep{Lambert1992} handles excess zeros in count data and could in principle be applied to integer percentages, but its support is unbounded above. In addition, some zeros reflect structural absence where ERC has not established, while others reflect low cover recorded as zero. Attributing all zeros to slow spread would bias the estimated diffusion and growth coefficients.

We propose a Bayesian hierarchical model that addresses both challenges. The model couples a zero-inflated binomial observation component, which accommodates the discrete 0–100 scale and separates structural absence from low observed cover, with an ecological diffusion–logistic growth PDE whose diffusion and growth coefficients depend on environmental covariates. Our contributions are twofold. First, the model represents the spread of tree cover across space and its local growth as separate processes and relates each to environmental covariates, using tree cover data alone. Second, we derive a homogenized equation for ecological diffusion with spatially varying diffusion and growth coefficients, which makes Bayesian inference feasible on satellite-resolution grids. We illustrate the approach with tree cover data from a subregion of the Niobrara River valley in the Nebraska Sandhills. In this subregion, elevation is strongly associated with where tree cover accumulates but only weakly with its local growth rate, whereas distance to the nearest stream is associated with both. 

The remainder of the paper is organized as follows. Section 2 describes the study region and data. Section 3 presents the observation and process models, and Section 4 describes Bayesian inference and the homogenization approach used to solve the PDE efficiently. Section 5 reports the analysis of the tree cover data, and Section 6 discusses findings, limitations, and extensions. Simulation studies, technical derivations, and MCMC diagnostics are provided in the Supplementary Material.

\section{Study Region and Data}\label{sec: sandhills}

\begin{figure}[htp]
  \centering
  \begin{subfigure}{0.68\textwidth}
    \centering
    \includegraphics[width=\linewidth]
      {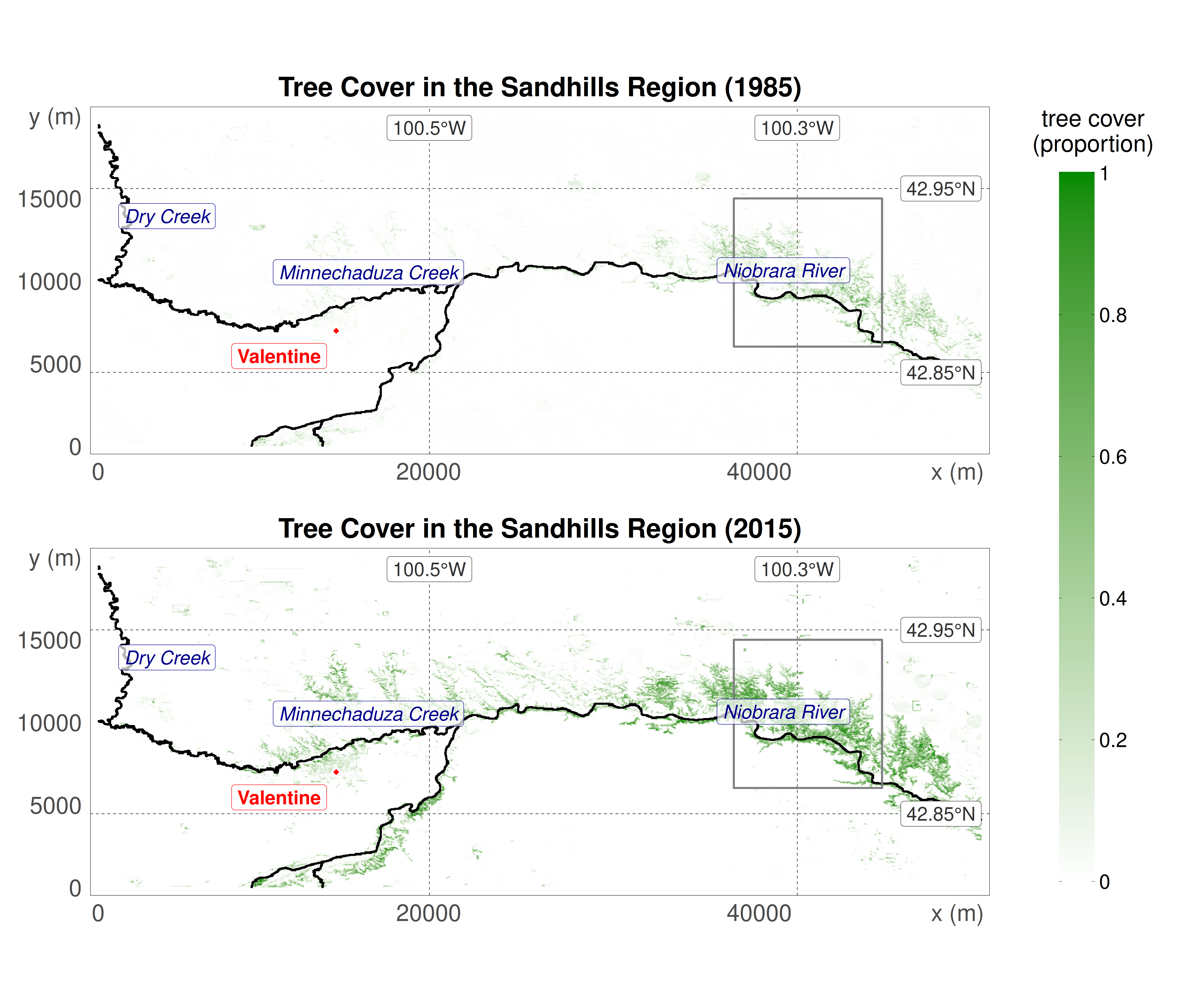}
    \caption{}\label{fig: obs_8515_real_data_full}
  \end{subfigure}
  \hfill
  \begin{subfigure}{0.68\textwidth}
    \centering
    \includegraphics[width=\linewidth]
      {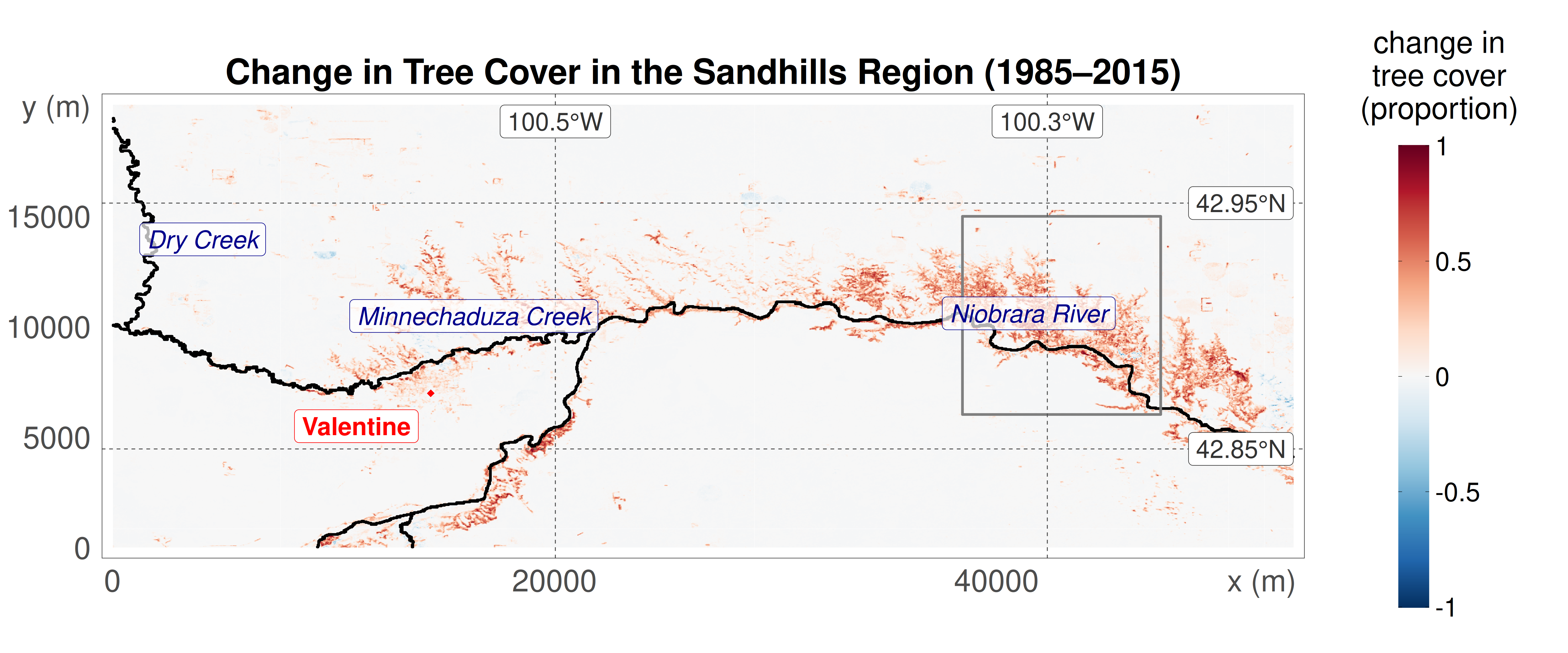}
    \caption{}\label{fig: obs_diff_8515_real_data_full}

    \centering
    \includegraphics[width=\linewidth]
      {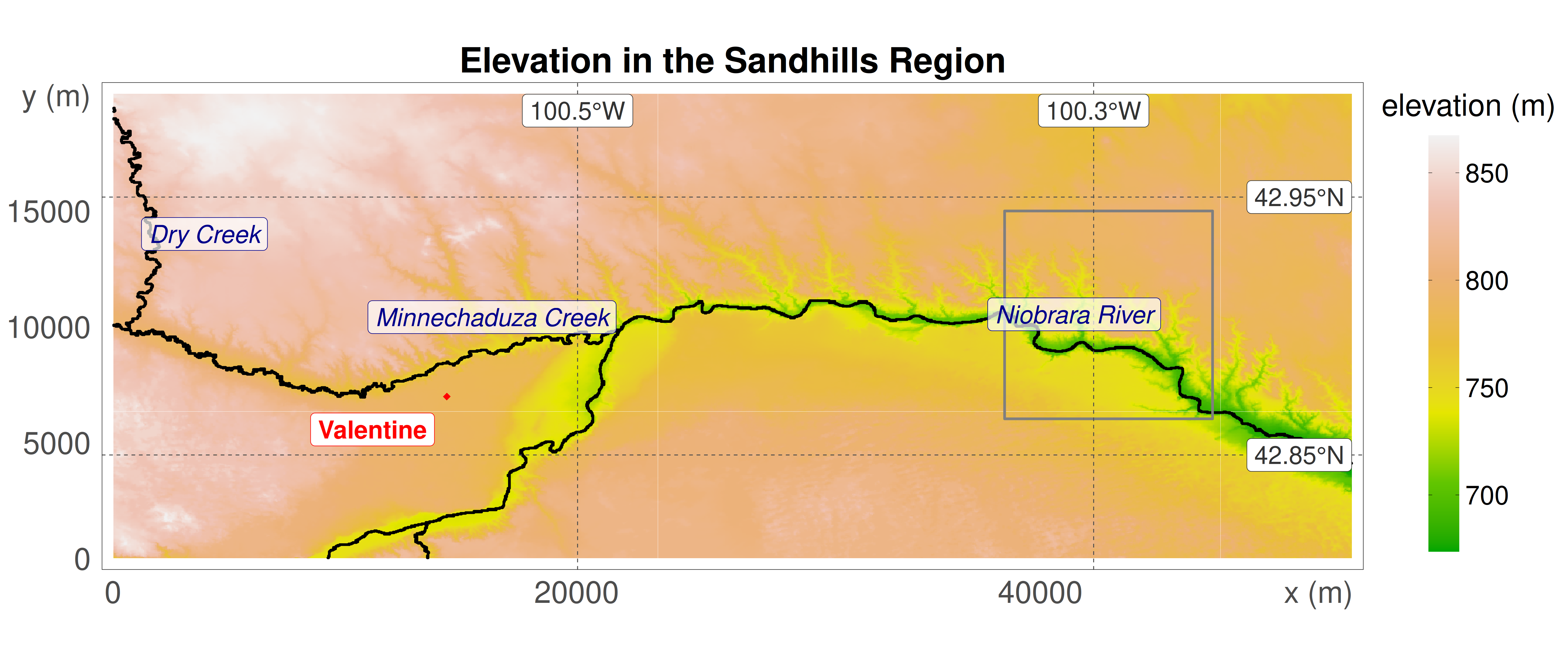}
    \caption{}\label{fig: elevation_real_data_full}
  \end{subfigure}
  \caption{Tree cover in the Sandhills region of Nebraska along the Niobrara River. 
  (a) Proportion of each pixel covered by trees in 1985 and 2015.
  (b) Change in tree cover between 1985 and 2015.
  (c) Elevation (m).
  Streams are shown by the black lines. The axes represent spatial coordinates measured from the southwest corner of the region. The gray box denotes the subregion used in data analysis.}
  \label{fig: real_data_full}
\end{figure}

We study a subregion of Cherry County, Nebraska, bounded by longitudes $[-100.335, -100.254]$ and latitudes $[42.864, 42.945]$ (gray box in Figure~\ref{fig: real_data_full}). The subregion lies within the Niobrara National Scenic River corridor \citep{NPS2006}, where limited cultivation and protected status make it well suited to studying encroachment with little confounding from land-use change.

We obtain annual tree cover estimates for 1985--2020 from the Rangeland Analysis Platform (RAP; \citealp{Allred2021}; \url{https://rangelands.app}), which estimates fractional cover of several plant functional types from Landsat imagery at 30~m resolution. The subregion is a $300 \times 300$ grid of RAP pixels, giving $N = 90{,}000$ pixels per year and an extent of about 9~km $\times$ 9~km. Tree cover is recorded as an integer percentage from 0 to 100. Many pixels have zero tree cover, particularly in early years, and some reach complete cover in later years (Figure~\ref{fig: obs_8515_real_data_full}). Tree cover in the broader region increased substantially between 1985 and 2015, most rapidly along streams and near existing trees (Figure~\ref{fig: obs_diff_8515_real_data_full}).

RAP does not distinguish tree species, so the recorded cover includes all trees and not only ERC. We use increases in tree cover as a proxy for ERC encroachment, because the subregion is predominantly grassland and increases in tree cover on grassland in this area can largely be attributed to ERC \citep{shrestha_current_2025}. Some of the recorded increase may nevertheless come from other tree species, and we return to this limitation in Section~6.

We use two environmental covariates, distance to the nearest stream and elevation. As shown in Figure~\ref{fig: real_data_full}, tree cover tends to be higher near streams and at lower elevations. The figures show elevation in meters. For model fitting, each covariate is scaled to $[0,1]$ over the subregion as $(x - x_{\min})/(x_{\max} - x_{\min})$, where $x$ is the value of that covariate at a pixel, either distance to the nearest stream or elevation, and $x_{\min}$ and $x_{\max}$ are its minimum and maximum over the subregion. A value of 0 therefore corresponds to the pixel nearest a stream or at the lowest elevation, and a value of 1 to the pixel farthest from the stream (6748.5~m) or at the highest elevation (125.2~m above the lowest).




\section{Methods}\label{sec:methods}

\subsection{Data Model}\label{sec.data_model}

 We propose a hierarchical model in which tree cover observations follow a zero-inflated binomial distribution whose success probability is the solution of the mechanistic PDE \eqref{pde_logis_growth} in Section~\ref{sec.PDE}. Because tree cover is recorded as an integer percentage (Section~2), with excess probability mass at 0 and a small but non-negligible concentration at 100, the zero-inflated binomial is a natural modeling choice. The zero-inflation component handles structural absences, and the binomial upper bound accommodates the observed mass at complete cover.

A random variable $Y$ following a zero-inflated binomial distribution,
$Y \sim \mathrm{ZIBinom}(\gamma, n, \mu)$, has the probability mass function,
\begin{equation}\nonumber
    f(y;\gamma, n, \mu) =  \begin{cases}  
        \gamma+ (1-\gamma) p(y=0; \,n, \mu), &  y = 0,\\
        (1-\gamma ) p(y; \,n, \mu), &  y = 1,\dots,n,
\end{cases}
\end{equation}
where $p(y; \,n, \mu)$ is the probability mass function of a binomial distribution with $n$ trials and success probability $\mu$, and
$\gamma \in (0,1)$ is the zero-inflation parameter representing
structural zeros.

The tree cover data consist of annual observations on an $n_1 \times n_2$ rectangular grid for years $\tau=1,\dots,T$. Year $\tau = 0$ supplies the initial condition for the process model in Section~\ref{sec.PDE}. Let $Y_{i,j,\tau}\in\{0,1,\dots,100\}$ denote the observed percent tree cover in pixel $(i,j)$ at year $\tau$, where $i=1,\dots,n_1$ and $j=1,\dots,n_2$, and $Y_{i,j,\tau}=100$ indicates complete cover. Conditional on $\gamma_{i,j,\tau}$ and $\mu_{i,j,\tau}$, the observations are assumed to be mutually independent with
\begin{equation}
    \label{equ: data model}
    Y_{i,j,\tau}\mid \gamma_{i,j,\tau},\mu_{i,j,\tau}\;\sim\;\text{ZIBinom}\left(\gamma_{i,j,\tau},\, n=100,\, \mu_{i,j,\tau}\right),
\end{equation}
where $\gamma_{i,j,\tau}$ captures the probability of structural absence of invasive tree cover, and the binomial success probability $\mu_{i,j,\tau} \in [0,1]$ governs the expected cover given the pixel is not structurally absent, and is linked to the mechanistic PDE process described in Section~\ref{sec.PDE}. Both parameters may in general vary over space and time. In this application we use only time-invariant environmental covariates, so $\gamma_{i,j,\tau}=\gamma_{i,j}$ for all $\tau$; the binomial success probability $\mu_{i,j,\tau}$ still varies in time through the evolution of the PDE solution. The binomial trial size is fixed at $n=100$ to match the integer percentage scale. This mixture formulation distinguishes structural absences, captured by $\gamma_{i,j}$, from zeros that arise when the mechanistic process predicts low cover (small $\mu_{i,j,\tau}$) and the observed values, subject to measurement error, happen to be zero.

Using observed tree cover as a proxy for ERC introduces uncertainty from multiple sources, including sensor limitations, image processing artifacts, and non-target woody vegetation contributing to the recorded cover values. Rather than modeling these sources separately, we treat them as absorbed into the binomial sampling variability around the PDE-predicted mean $\mu_{i,j,\tau}$, acknowledging that the recorded pixel value reflects true ERC cover plus unattributable background noise.

Structural zeros in tree cover are often associated with environmental constraints. We therefore model $\gamma_{i,j}$ as a function of environmental covariates. Let $\bX_{i,j}=(X_{0,i,j},X_{1,i,j},\ldots,X_{p,i,j})^\top$ denote the $(p+1)$-dimensional covariate vector for pixel $(i,j)$, where $X_{0,i,j}=1$ is the intercept. The zero-inflation parameter is modeled via a logistic link, $\gamma_{i,j}=\left\{1+\exp\left(-\bX_{i,j}^\top\bbeta_{\gamma}\right)\right\}^{-1},
$ where $\bbeta_{\gamma}$ is a $(p+1)$-dimensional vector of unknown coefficients, and the link function ensures that $\gamma_{i,j} \in (0,1)$.

\subsection{Mechanistic Process Model}\label{sec.PDE}

ERC spreads mainly by seed \citep{UNLBEEF2025}. New trees therefore tend to appear near existing ones, and whether they persist depends on local conditions. Once trees are present at a location, cover increases through infilling and canopy expansion, independent of dispersal from neighboring areas. We represent the first process by ecological (Fokker–Planck) diffusion with a spatially varying diffusion coefficient and the second by logistic growth  \citep{Acevedo2012}. Because we observe tree cover rather than seeds or seedlings, the diffusion coefficient describes the net outcome of dispersal and establishment, and we do not attempt to separate the two.

To capture the spatiotemporal dynamics of invasive trees and their 
dependence on environmental factors, we model the binomial 
success probability $\mu_{i,j,\tau}$ by the value of a continuous 
process $\mu(\bs,t)$ evaluated at the center of pixel $(i,j)$ at 
time $\tau$, where $\bs=(s_1,s_2)^\top\in\Omega$ 
and $t\in[0,\infty)$, and $\Omega\subset\mathbb{R}^2$ is the bounded rectangular domain corresponding to the $n_1\times n_2$ observation grid. This is a reasonable approximation because tree 
cover varies gradually relative to the $30\,\text{m}\times 30\,\text{m}$ 
pixel size, so the continuous process value at the pixel center is representative of the pixel average. The process $\mu(\bs,t)$ is governed by a PDE with ecological diffusion and logistic growth,
\begin{equation}\label{pde_logis_growth}
        \frac{\partial }{\partial t}\mu(\bs,t)=\nabla^2_{\bs}\!\bigl[D(\bs)\,\mu(\bs,t)\bigr]+G(\bs)\mu(\bs,t)\left(1-\frac{\mu(\bs,t)}{K(\bs)}\right),
\end{equation}
subject to a given initial condition $\mu(\bs, 0)$ for $\bs \in \Omega$, which in our application is the observed tree cover in the first year, and the boundary condition $\mu(\bs, t)=0$ for all boundary locations $\bs \in \partial\Omega$.
Here $\nabla^2_{\bs}=\frac{\partial^2}{\partial s_1^2}+\frac{\partial^2}{\partial s_2^2}$ denotes the Laplacian with respect to $\bs$. The spatially varying quantities $D(\bs)$, $G(\bs)$, and $K(\bs)$ represent the diffusion coefficient, growth coefficient, and carrying capacity, respectively. 

The first term in \eqref{pde_logis_growth}, $\nabla^2_{\bs}\bigl[D(\bs)\,\mu(\bs,t)\bigr]$, models ecological diffusion. When this term is positive, the focal location has lower tree cover (and thus a smaller seed source) than its neighbors, leading to an influx from surrounding areas and an expected increase in local cover. Conversely, a negative value suggests the location has higher tree cover than its surroundings, resulting in a net outflow of trees to adjacent sites. 

The diffusion coefficient $D(\bs)$ describes how readily tree cover that spreads from neighboring locations takes hold at location $\bs$. Where $D(\bs)$ is small, new cover tends to take hold and accumulate. Where $D(\bs)$ is large, little new cover is retained, and tree cover stays sparse. Environmental factors, such as proximity to streams and elevation, can influence diffusion rates. To capture this dependence, we model $D(\bs)$ as a function of environmental covariates,
$D(\bs)=\exp\left(-\bX(\bs)^\top \bbeta_D\right),$
where $\bX(\bs)$ is a $(p+1)$-dimensional covariate vector and $\bbeta_D$ is a $(p+1)$-dimensional vector of unknown parameters to be estimated from the data. The exponential form ensures $D(\bs) > 0$ for all $\bs$, as required for a valid diffusion coefficient.

 An alternative is Fickian diffusion, $\frac{\partial }{\partial t} \mu(\bs,t) = \nabla_{\bs}\cdot\bigl[D(\bs)\nabla_{\bs}\mu(\bs,t)\bigr]$, where $\nabla_{\bs}=\left(\frac{\partial}{\partial s_1}, \frac{\partial}{\partial s_2}\right)^\top$ denotes the spatial gradient operator with respect to $\bs$. Fickian diffusion assumes that spread is proportional to the gradient of tree cover, so cover spreads from areas of high cover toward areas of low cover at a rate governed by $D(\bs)$, with no intrinsic directional preference \citep{Andreucci2019}. With zero-flux boundary conditions and no growth, the Fickian steady state is spatially constant. Tree cover would eventually equalize across the landscape, and spatial variation in $D(\bs)$ would affect only the rate of spread, not the final distribution. Fickian diffusion is therefore difficult to interpret ecologically, because the resulting spatial pattern does not depend on local environmental conditions \citep{Bengfort2016}. By contrast, we employ ecological diffusion, an environmentally mediated diffusion formulation, which is appropriate when spatial heterogeneity arises from local environmental conditions that affect how readily tree cover takes hold \citep{Bengfort2016, Bringuier2011, Garlick2011}. Under ecological diffusion, with zero-flux boundary conditions and no growth, the steady-state solution is proportional to $1/D(\bs)$, so tree cover accumulates where $D(\bs)$ is small. This behavior is consistent with the observed pattern that tree cover establishes preferentially in certain areas.

The second term in \eqref{pde_logis_growth}, $G(\bs)\mu(\bs,t)\left(1-{\mu(\bs,t)}/{K(\bs)}\right)$, represents logistic growth and models the temporal increase in tree cover at each location. Here, $G(\bs)$ denotes the local growth rate, with higher values corresponding to more rapid growth toward the carrying capacity. Declines in tree cover are rare and small in this region, so we restrict $G(\bs)$ to be positive. The carrying capacity $K(\bs)$ defines the upper bound for tree cover; since tree cover is measured as a proportion and cannot exceed $100\%$, we set $K(\bs)=1$ for all locations. This ensures that $\mu(\bs,t)$ remains in the interval $[0,1]$, providing a biologically meaningful and valid probability for the binomial observation model. The growth coefficient $G(\bs)$ may also depend on environmental covariates and is modeled similarly to the diffusion coefficient,
$G(\bs)=\exp\left(-\bX(\bs)^\top\bbeta_G\right),$
where $\bbeta_G$ is a $(p+1)$-dimensional vector of unknown parameters to be estimated from the data.

\section{Bayesian Inference and Computation}\label{posterior}

\subsection{Bayesian Inference}
Inference for the model parameters is performed within a Bayesian framework using Markov chain Monte Carlo (MCMC) methods.
Let $ \bY $ denote the vector of all observations $Y_{i,j,\tau}$ stacked over $i,j$ and $\tau$. The joint posterior distribution of the parameter vectors is proportional to the product of the likelihood and the prior distributions, $f(\bbeta_D, \bbeta_G, \bbeta_{\gamma} \mid \bY) \propto L( \bY \mid \bbeta_D, \bbeta_G, \bbeta_{\gamma}) \, \pi(\bbeta_D)\,\pi(\bbeta_G)\,\pi(\bbeta_{\gamma})$. To complete the prior specification, we assign independent, weakly informative multivariate normal priors, $\bbeta_a \sim N\left(\mathbf{0},\,\Sigma_a\right)$ for $a\in\{D,G,\gamma\}$, to each of the parameter vectors governing diffusion, growth and zero-inflation, where each $\Sigma_a$ is diagonal with large variance to reflect weak prior information. Closed-form full conditionals are unavailable; therefore, we update each parameter block using the Metropolis--Hastings algorithm.

Each likelihood evaluation requires solving the PDE in \eqref{pde_logis_growth}. The PDE is solved numerically from the initial condition to each observation time, using internal time steps $\Delta t$ chosen for numerical stability (typically $\Delta t < 1$ year); the computed solution at observation times $\tau = 1,\dots,T$ yields the values $\mu_{i,j,\tau}$ needed for the likelihood. Implementation details are provided in Section~\ref{sec:solvepde}. Given $\mu_{i,j,\tau}$, the likelihood function $L(\bY \mid \bbeta_D, \bbeta_G, \bbeta_{\gamma})$ under the zero-inflated binomial observation model in \eqref{equ: data model} is
\begin{equation}\nonumber
\begin{split}
L(\bY \mid \bbeta_D, \bbeta_G, \bbeta_\gamma)
&=\prod_{i=1}^{n_1}\prod_{j=1}^{n_2}\prod_{\tau=1}^{T}
\Bigg[
\mathds{1}\{Y_{i,j,\tau}=0\}\Big\{\gamma_{i,j,\tau}+(1-\gamma_{i,j,\tau})(1-\mu_{i,j,\tau})^{100}\Big\}\\
&\qquad\qquad+\mathds{1}\{Y_{i,j,\tau}>0\}(1-\gamma_{i,j,\tau})
\binom{100}{Y_{i,j,\tau}}(\mu_{i,j,\tau})^{\,Y_{i,j,\tau}}(1-\mu_{i,j,\tau})^{\,100-Y_{i,j,\tau}}
\Bigg].
\end{split}
\end{equation}

\subsection{Solving the PDE using Homogenization}\label{sec:solvepde}
Our PDE model in \eqref{pde_logis_growth} does not have a closed-form solution; therefore, we solve it numerically using the finite-difference method \citep{Haberman2013}. This approach discretizes the spatial domain into $N$ grid points and the temporal domain into $M$ steps of length $\Delta t$; accuracy improves as $N$ increases and $\Delta t$ decreases. The computational cost of the explicit finite-difference scheme is $\mathcal{O}(NM)$ per MCMC iteration, since each grid point is updated once per time step. To obtain $\mu(\bs,t)$ at the observational grid and times, we may take $N=n_1\times n_2$ and $M=T$, or use a finer spatial and/or temporal resolution if numerical errors are substantial (e.g., the solution falls outside $[0,1]$). However, this cost becomes prohibitive for high-resolution grids such as ours, because Bayesian inference via MCMC requires evaluating $\mu_{i,j,\tau}$ at every iteration.

We address this computational bottleneck using homogenization \citep{Holmes2013}, which has been applied to ecological diffusion models \citep{Powell&Zimmermann2004,Garlick2011} and embedded in Bayesian hierarchical models \citep{Hefley2017a,Lu2019}. Applying this approach requires first deriving the homogenized PDE associated with the specific equation under consideration. In this work, we derive the homogenized PDE for the proposed ecological diffusion model with logistic growth and spatially varying diffusion and growth coefficients. \citet{Lu2019} derived a related homogenized PDE for spatially varying diffusion with a constant growth rate. Full derivations are provided in Supplementary Material~S.1; 
below we summarize the main steps. 

The key idea is to solve the PDE on a coarser spatial scale by applying a change of variables. Let $\Omega$ denote the fine-resolution spatial domain and $\Omega_c$ the corresponding coarse-resolution domain. We introduce the coarse-scale spatial coordinate $\br$ through the transformation $\br = \epsilon \bs$, where $0 < \epsilon < 1$ represents the ratio between the coarse and fine resolutions. For example, if $\bs$ is measured in meters and $\br$ in kilometers, then $\epsilon = 1/1000$. Instead of solving the PDE directly at the fine resolution, the homogenization framework yields an approximate PDE defined at the coarse scale whose solution captures the aggregate behavior of the fine-scale dynamics. In practice, the spatially varying diffusion and growth coefficients are averaged over coarse-grid cells, allowing the PDE to be solved on a reduced-resolution domain and substantially reducing computational cost \citep{Pavliotis&Stuart2008}. This approximation works well when the spatially heterogeneous coefficients vary rapidly relative to the macroscopic spatial structure of interest, so that fine-scale fluctuations are effectively averaged out in the homogenized equation. In such cases, the scale ratio $\epsilon$ is determined by the ratio between the scale of the substructure variation and the larger-scale spatial structure of interest.

The homogenized PDE that approximates \eqref{pde_logis_growth} is given below. Let $c(\br,t)$ denote the transformed process at the coarse scale. Then, $c(\br,t)$ satisfies
\begin{equation}\label{homog_heateq}
    \frac{\partial }{\partial t}c(\br,t)=\overline{D}(\br)\nabla^2_{\br}c(\br,t)+\overline{G_1}(\br)c(\br,t)-\overline{G_2}(\br)c(\br,t)^2,
\end{equation}
subject to the initial condition $c(\br, 0)=\frac{\int_{\mathcal{A}} D(\br,\bs)\mu(\br,\bs,0)\;d \bs}{\int_{\mathcal{A}} 1 \;d \bs}$, $\br \in \Omega_c$, and the boundary condition $c(\br, t)=0$, $\br \in \partial\Omega_c$. The homogenized coefficients are given by
\begin{equation}\label{equ:homog_coef}
    \overline{D}(\br)=\frac{\int_{\mathcal{A}}1 \;d \bs}{\int_{\mathcal{A}}  \frac{1}{D(\br,\bs)} \;d \bs}, \quad
    \overline{G_1}(\br)=\frac{\int_{\mathcal{A}} \frac{G(\br,\bs)}{D(\br,\bs)} \;d\bs}{\int_{\mathcal{A}}  \frac{1}{D(\br,\bs)} \;d \bs}, \quad
    \overline{G_2}(\br)=\frac{\int_{\mathcal{A}} \frac{G(\br,\bs)}{D^2(\br,\bs)} \;d\bs}{\int_{\mathcal{A}}  \frac{1}{D(\br,\bs)} \;d \bs},
\end{equation}
where $\mathcal{A}$ is the spatial aggregation area of the coarse-grid cell at $\br$, that is, the region over which fine-scale values are averaged to produce the value at that coarse-grid point. Here $D(\br,\bs)$, $G(\br,\bs)$, and $\mu(\br,\bs,0)$ denote $D(\bs)$, $G(\bs)$, and $\mu(\bs,0)$ at the fine-scale locations $\bs$ within that cell. The homogenized PDE can be solved numerically on a much coarser grid of size $N_c$, where $N_c = \epsilon^2 N$, which greatly reduces computational cost. Once the homogenized solution $c(\br,t)$ is obtained, the fine-scale solution of the original PDE can be approximated by 
$\mu(\bs,t) \approx c(\br,t)/D(\bs)$, where $c(\br,t)$ is the coarse-grid solution evaluated at the coarse-grid location $\br$ corresponding to the grid element containing $\bs$. Detailed numerical implementations for solving both the homogenized PDE in \eqref{homog_heateq} and the original PDE in \eqref{pde_logis_growth} are provided in Supplementary Material~S.2.

The full implementation of Bayesian inference via MCMC is summarized in Supplementary Material~S.3. 
The sampler is implemented in \texttt{R}, with computationally intensive components (homogenized-coefficient aggregation and the finite-difference PDE solver) coded in \texttt{Rcpp} to reduce runtime. Proposal distributions are tuned to target acceptance rates of approximately $40\%$ for block updates of $\bbeta_D$, $\bbeta_G$, and $\bbeta_{\gamma}$. MCMC convergence is assessed using multiple chains, trace plots, effective sample sizes, and the Gelman--Rubin diagnostic. Model fit is evaluated via posterior predictive checks.

\section{Application to the Sandhills}\label{sec: data_analysis}

Before applying the proposed framework to the tree cover data, we validated it through simulation studies under two settings, one with simplified synthetic covariates and one using observed covariates from the study region. The PDE-based method recovers the true parameters, with 95\% credible intervals achieving near-nominal coverage, and outperforms competing approaches in predictive coverage and test-set mean squared error (MSE). Full details are provided in Supplementary Material~S.4. 

\paragraph{Data Analysis.}

We analyze the $300 \times 300$ grid of 30~m pixels described in Section~2, with annual tree cover from 1985 ($\tau = 0$) to 2020 ($\tau = 35$). The covariate vector includes an intercept, scaled distance to the nearest stream, and scaled elevation. We use data from 1985--2015 as the training set and reserve 2016--2020 as a held-out test set. The observation $Y_{i,j,0}$ in year $1985$, divided by 100 to convert it to a proportion, serves as the initial condition $\mu_{i,j,0}$. We assign weakly informative priors with covariance matrix $\Sigma_a = 100^2\mathbf{I}$ for each $a\in\{D,G,\gamma\}$. We solve the homogenized PDE~\eqref{homog_heateq} with $\epsilon=1/15$, which aggregates each $15 \times 15$ block of pixels into one coarse cell of 450~m $\times$ 450~m and reduces the grid size to $20\times 20$. This choice is justified by the spatial structure of elevation in the study region: while elevation may vary locally within individual pixels, its large-scale spatial trend varies slowly relative to the $\sim9\,\text{km}$ extent of the study domain, satisfying the separation of scales required for homogenization, so that averaging the diffusion and growth coefficients over coarse-grid cells incurs no significant loss of large-scale ecological information. We run three independent chains from distinct starting values for 20,000 Metropolis--Hastings iterations each, discard the first 5,000 as burn-in, and thin the remainder to 1,000 posterior samples. All three chains converge, and full diagnostics are reported in Supplementary Material~S.5.

We compare the proposed PDE-based model with three benchmarks, a generalized additive model (GAM), a spatial binomial hierarchical model fitted via integrated nested Laplace approximation (INLA), and a regression tree (RT). The GAM includes smooth terms for elevation, time, and distance to the nearest stream, along with a two-dimensional thin-plate spline over the spatial coordinates. The INLA model uses a binomial likelihood with a linear predictor comprising fixed effects for distance to the nearest stream, elevation, time, and spatial coordinates; residual spatial dependence is captured by an intrinsic conditional autoregressive (ICAR) random effect on the grid adjacency graph. The RT model serves as a nonparametric benchmark, with distance to the nearest stream, elevation, time, and spatial coordinates as predictors. All models are implemented in \texttt{R}, using the \texttt{mgcv}, \texttt{R-INLA} \citep{inla}, and \texttt{rpart} packages for the GAM, INLA, and RT, respectively.

\paragraph{PDE Parameter Estimates.}
Posterior means and 95\% credible intervals for the PDE-based model are summarized in Table~\ref{tab:pde_est_real}. All credible intervals exclude zero, but the effects differ greatly in size. Because both covariates are scaled to $[0,1]$, each coefficient compares the two ends of a covariate's range in the subregion, holding the other covariate fixed. For distance, these are the pixel nearest a stream and the pixel farthest from one (6748.5~m farther away). For elevation, they are the lowest and the highest points (125.2~m apart). From the nearest pixel to the farthest, $D(\bs)$ increases by a factor of $\exp(0.701) \approx 2.02$ and $G(\bs)$ decreases by a factor of $\exp(-0.424) \approx 0.65$. From the lowest to the highest elevation, $D(\bs)$ increases by a factor of $\exp(1.589) \approx 4.90$, whereas $G(\bs)$ changes little ($\exp(-0.044) \approx 0.96$). Under the homogenized model, $\mu(\bs,t) \approx c(\mathbf{r},t)/D(\bs)$ (Section~\ref{sec:solvepde}). At any time, expected cover within a coarse cell is proportional to $1/D(\bs)$, so $1/D(\bs)$ acts as a weight on where tree cover accumulates. This weight is about half as large at the farthest pixel as at the nearest pixel, and about one-fifth as large at the highest elevation as at the lowest. For a pixel nearest a stream and at the lowest elevation, the local growth rate is $\exp(-2.788) \approx 0.06$ per year, which corresponds to a doubling time of about 11 years at low cover. Elevation is thus strongly associated with where tree cover accumulates but only weakly with how fast it grows, whereas distance to the nearest stream is associated with both. 

The zero-inflation coefficients show that structural absence depends on both covariates, in opposite directions, and more strongly on elevation. Because $\gamma$ is modeled with a logistic link, the exponentiated coefficients are odds ratios for structural absence, where the odds are $\gamma/(1-\gamma)$. For a pixel nearest a stream and at the lowest elevation, the probability of structural absence is $\{1+\exp(1.805)\}^{-1} \approx 0.14$. Holding distance fixed, the odds are $\exp(1.286) \approx 3.6$ times higher at the highest elevation than at the lowest, which raises this probability to about 0.37 and is consistent with uplands that remain grassland. Holding elevation fixed, the odds at the farthest pixel are $\exp(-0.759) \approx 0.47$ times those at the nearest pixel, which lowers the probability to about 0.07. This difference is small compared with that for elevation.

\begin{table}[H]
\centering
\caption{Posterior means and 95\% credible intervals for the parameters of the PDE-based model fitted to the tree cover data from the study subregion. Covariates are scaled to $[0,1]$, so each slope compares the two ends of a covariate's range. Because $D(\bs) = \exp\{-X(\bs)^\top{\bbeta}_D\}$ and $G(\bs) = \exp\{-X(\bs)^\top{\bbeta}_G\}$, a negative coefficient means that the quantity increases with the covariate.}
\label{tab:pde_est_real}
    \begin{tabular}{|c|c|c|}
        \hline
        \textbf{Parameter} & \textbf{Estimate} & \textbf{$95\%$ Credible Interval}\\
        \hline
        $\beta_{D,\,\text{intercept}}$ & $-5.230$ & $(-5.237, -5.222)$\\
        \hline
        $\beta_{D,\text{dist}}$ & $-0.701$ & $(-0.719, -0.684)$\\
        \hline
        $\beta_{D,\text{elev}}$ & $-1.589$ & $(-1.591, -1.586)$\\
        \hline
        $\beta_{G,\,\text{intercept}}$ & 2.788 & (2.786, 2.790)\\
        \hline
        $\beta_{G,\text{dist}}$ & 0.424 & (0.418, 0.430)\\
        \hline
        $\beta_{G,\text{elev}}$ & 0.044 & (0.040, 0.049)\\
        \hline
        $\beta_{\gamma,\,\text{intercept}}$ & $-1.805$ & $(-1.818, -1.789)$\\
        \hline
        $\beta_{\gamma,\text{dist}}$ & $-0.759$ & $(-0.788, -0.731)$\\
        \hline
        $\beta_{\gamma,\text{elev}}$ & 1.286 & (1.259, 1.312)\\
        \hline
    \end{tabular}
\end{table}



\begin{figure}[H]
    \centering
    \includegraphics[width=1.0\linewidth]{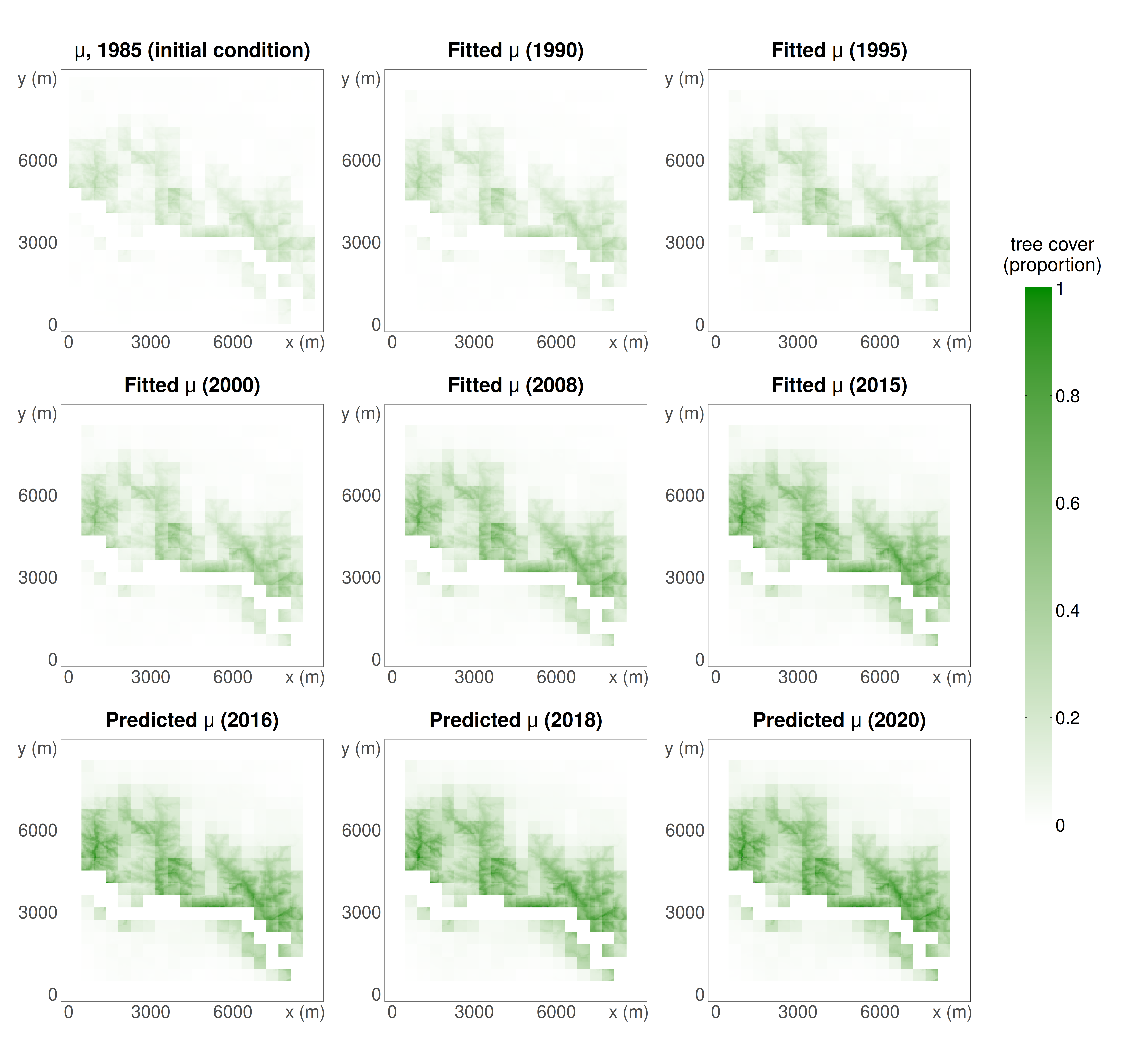}
    \caption{PDE solution $\mu(\bs,t)$, the expected proportion of tree cover before the zero-inflation component is applied, obtained by solving the homogenized PDE at the posterior means of the parameters. The top two rows show training years (1985, 1990, 1995, 2000, 2008, and 2015), where the values are fitted, and the bottom row shows test years (2016, 2018, and 2020), where they are predicted. The 1985 panel shows the homogenized initial condition, which differs from the observed 1985 cover.}
    \label{pic: mu_real_data}
\end{figure}

Figure~\ref{pic: mu_real_data} shows the PDE solution $\mu(\bs,t)$ over time. Tree cover gradually accumulates near the Niobrara River corridor and at lower elevations, then spreads outward to neighboring areas over time, consistent with the diffusion--growth dynamics of ERC encroachment captured by the mechanistic model.

\begin{figure}[H]
    \centering
    \includegraphics[width=\linewidth]{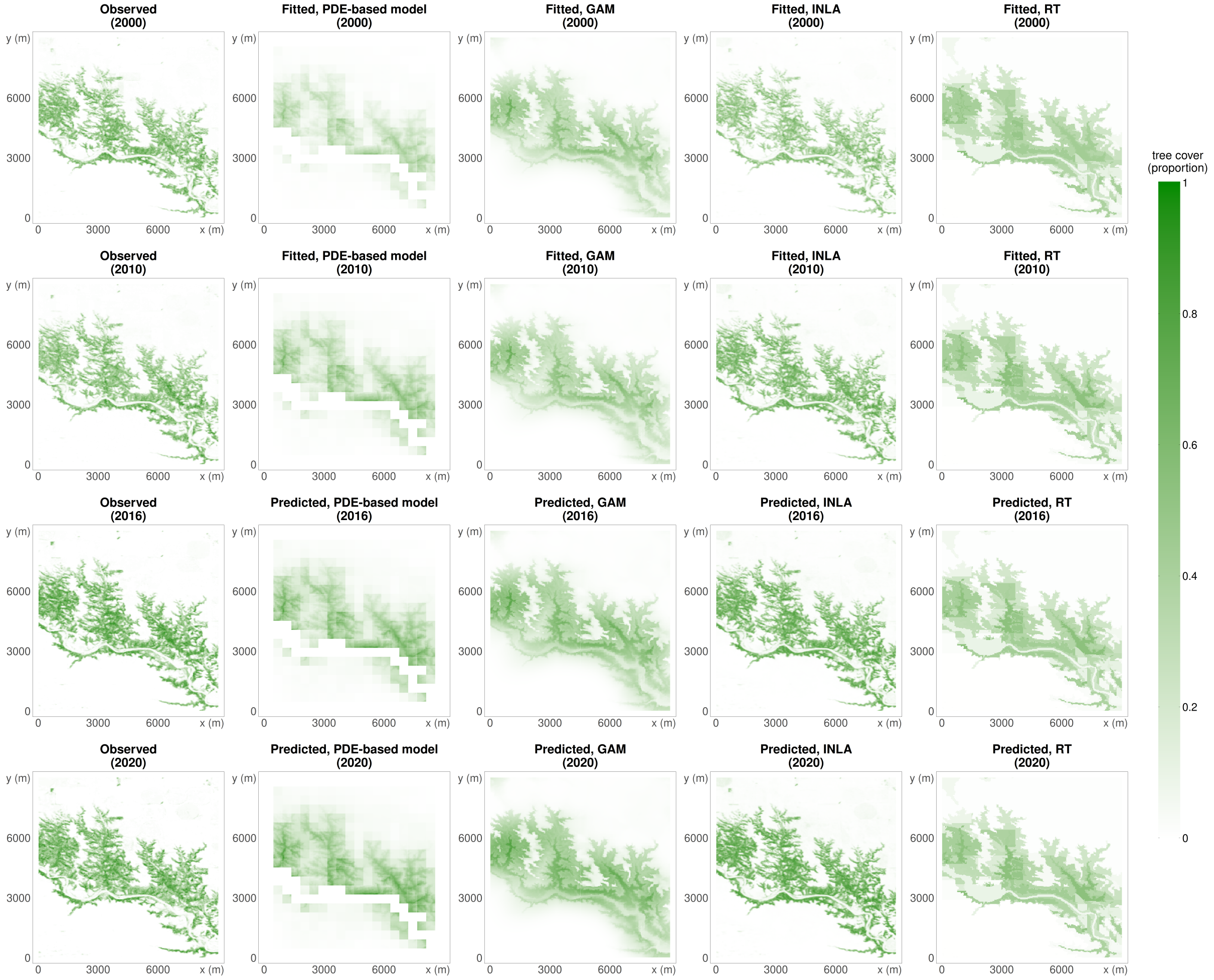}
    \caption{Observed proportion of tree cover (first column) and the fitted and predicted proportions from the PDE-based model (second column), GAM (third), INLA (fourth), and regression tree (fifth). Values are fitted in the training years 2000 and 2010 and predicted in the test years 2016 and 2020. For the PDE-based model, they equal $(1-\gamma_{i,j})\mu_{i,j,\tau}$, which combines the PDE solution in Figure~\ref{pic: mu_real_data} with the zero-inflation component. The blocky structure in the PDE column reflects the coarse grid used for homogenization.}
    \label{pic: dist_real_data}
\end{figure}

\paragraph{Model Comparison.}
Spatial maps of the fitted and predicted tree cover from all four methods at selected years are shown in Figure~\ref{pic: dist_real_data}. All methods reproduce the broad spatial pattern. The PDE-based method captures the dominant large-scale spatial dynamics of ERC encroachment; fine-scale spatial detail is somewhat attenuated relative to the observations, a consequence of the homogenization procedure that averages the diffusion and growth coefficients over coarse-grid cells.  Among the benchmark methods, INLA performs most closely to the PDE-based method in reproducing spatial patterns. The GAM systematically underestimates tree cover, and RT predictions exhibit blocky spatial structure.

Table~\ref{tab:compare} compares the four methods. INLA achieves the lowest MSE on both training and test data, while the PDE-based model and RT perform comparably, and the GAM performs worst. However, the 95\% posterior predictive coverage of the PDE-based model (71.9\% training, 77.8\% test) substantially exceeds that of all three comparators, none of which exceeds 27\% on either dataset, indicating that the PDE-based model provides better-calibrated uncertainty quantification while the alternative methods are severely overconfident.

\begin{table}[h]
    \centering
    \caption{MSE on the proportion scale and 95\% posterior predictive coverage for all four methods on both training and test data.}
    \label{tab:compare}
    \begin{tabular}{|c|c|c|c|c|}
        \hline
          & \textbf{MSE} & \textbf{MSE} & \textbf{Pred. Coverage} & \textbf{Pred. Coverage}\\
        \textbf{Method} & \textbf{(Training Data)} & \textbf{(Test Data)} & \textbf{(Training Data)} & \textbf{(Test Data)}\\
        \hline
        PDE-Based & 0.013 & 0.019 & 71.9\% & 77.8\%\\
        \hline
        GAM & 0.015 & 0.022 & 4.2\% & 2.6\%\\
        \hline
        INLA & 0.003 & 0.004 & 26.4\% & 21.0\%\\
        \hline
        RT & 0.011 & 0.016 & 19.4\% & 10.7\%\\
        \hline
    \end{tabular}
\end{table}

All four methods agree that tree cover is higher near streams and at lower elevations. Of the four, only the PDE-based and INLA models summarize covariate effects as scalar coefficients, so we compare parameter estimates for these two. Fitted to the same tree cover data, the INLA model gives fixed-effect estimates on the log-odds scale of $-4.589$ $(-5.956, -3.222)$ for distance to the nearest stream, $-3.011$ $(-3.160, -2.861)$ for elevation, and $0.048$ $(0.047, 0.049)$ for time, with 95\% credible intervals in parentheses. The time effect corresponds to an increase of about 5\% in the odds of tree cover per year. These directions agree with the PDE-based estimates.

The two models differ, however, in what their parameters represent. INLA's coefficient for distance to stream is a spatially constant coefficient that describes the average association between stream proximity and tree cover across the entire landscape, with residual spatial structure absorbed by the ICAR random effect. The PDE-based model instead decomposes this association into two distinct ecological processes: a diffusion coefficient $D(\bs)$ governing the rate of dispersal from neighboring occupied locations and establishment, and a growth coefficient $G(\bs)$ governing the rate of local canopy expansion. Consequently, the PDE-based model not only identifies that stream proximity is associated with higher cover, as INLA does, but also indicates how the association arises. This decomposition may inform management, because limiting where new trees take hold and slowing the growth of existing trees call for different interventions.

Covariate interpretations for the GAM and RT are not pursued; both methods show deficiencies in prediction (Figure~\ref{pic: dist_real_data}) and low coverage (Table~\ref{tab:compare}), and neither produces scalar covariate summaries amenable to ecological interpretation, as the GAM represents effects as nonparametric smooth functions and the RT as threshold-based splitting rules.

\section{Discussion}
We have proposed a Bayesian mechanistic framework that combines an ecological diffusion--logistic growth PDE, a zero-inflated binomial observation model, and a homogenization-based computational strategy. The framework describes tree cover dynamics, their dependence on environmental covariates, and structural tree absence. In a subregion of the Niobrara River valley in the Nebraska Sandhills, the fitted model indicates that elevation is strongly associated with where tree cover accumulates but only weakly with its local growth rate, whereas distance to the nearest stream is associated with both. More generally, the framework is most useful when the scientific question concerns how a process unfolds, such as how fast tree cover increases and which environmental conditions favor its accumulation.

The analysis has limitations that suggest directions for further work. First, it is based on a single subregion of about 9~km $\times$ 9~km, so the estimated covariate effects describe this area and would need to be re-estimated elsewhere. Second, homogenization smooths spatial detail below the scale of a coarse cell, which is the cost of the computational gain. Third, the predictive coverage of the PDE-based model (72--78\%) is closer to the nominal level than that of the benchmarks but remains below it. We hypothesize that the shortfall reflects variation in the satellite-derived cover that the binomial observation model does not fully capture. Finally, tree cover is a proxy for ERC, because RAP does not distinguish tree species.

The framework can be extended in several directions. The diffusion and growth coefficients could depend on time-varying covariates such as fire history. The process model could include multiple interacting species, and faster numerical solvers would allow larger domains and longer time horizons. Formal model-selection criteria would support comparison among competing model structures.

\paragraph{Supplementary material.} The supplementary material includes the derivation of the homogenized PDE, the numerical schemes for solving both the original and homogenized PDEs, the MCMC algorithm, simulation studies under two settings, and MCMC diagnostics for the Sandhills application. Code and example data are available at \url{https://github.com/cj020/Spatio-Temporal_Dynamics_in_Invasive_Tree}.

\bibliographystyle{rss}
\bibliography{reference}
\end{document}